\documentclass[preprint2]{aastex}

\slugcomment{to appear in Ap. J.}

\shorttitle{Heating of the Hot IGM by Radio Galaxies and Gamma-Ray Emission}
\shortauthors{S. Inoue and S. Sasaki}

\begin{document}

\title{Heating of the Hot Intergalactic Medium by Powerful Radio Galaxies
       and Associated High Energy Gamma-Ray Emission}

\author{Susumu Inoue\altaffilmark{1,2}}
\email{inoue@th.nao.ac.jp}

\and

\author{Shin Sasaki\altaffilmark{3}}
\email{sasaki@phys.metro-u.ac.jp}

\altaffiltext{1}{Division of Theoretical Astrophysics,
                 National Astronomical Observatory,
                 2-21-1 Ohsawa, Mitaka, Tokyo, Japan 181-8588}
\altaffiltext{2}{Osservatorio Astrofisico di Arcetri,
                 Largo E. Fermi 5, I-50127, Firenze, Italy}
\altaffiltext{3}{Department of Physics, Tokyo Metropolitan University,
                 1-1 Minami-Ohsawa, Hachioji, Tokyo, Japan 192-0397}

\begin{abstract}

There is increasing evidence that some heating mechanism
 in addition to gravitational shock heating
 has been important
 for the hot gas inside clusters and groups of galaxies,
 as indicated by their observed X-ray scaling properties.
While supernovae are the most obvious candidate heating sources,
 a number of recent studies have suggested that
 they may be energetically insufficient.
Here we consider high-power, FRII radio galaxies
 and shock heating of the intracluster medium
 (ICM, including the case of the intergalactic medium prior to cluster formation)
 by their large-scale jets.
Based on the observed statistics of radio galaxies in clusters
 and their evolution, along with the most reasonable assumptions,
 it is shown that they can provide the ICM
 with excess specific energies of
 1--2 keV per particle,
 mainly during the redshift interval $z \sim 1-3$.
This naturally meets the requirements
 of cluster evolution models with non-gravitational feedback
 in accounting for the observed deviations in the X-ray luminosity-temperature relation.
In contrast to supernovae, such large-scale jets deposit their energy
 directly into the low density ICM outside galaxies,
 and are much less susceptible to radiative losses.
As a clear and potentially decisive test of this scenario,
 we propose the observation of `prompt' high energy gamma-rays
 emitted by shock-accelerated, non-thermal electrons
 during the epoch of ICM heating by radio galaxies,
 which may be feasible with the {\it GLAST} satellite.
Implications for recent detections of excess hard X-rays from groups
 are also discussed.

\end{abstract}


\section{Introduction}
 \label{sec:intro}

Within the framework of currently popular structure formation theories,
 the hot, X-ray emitting intracluster medium (ICM)
 observed in clusters and groups of galaxies
 (hereafter collectively referred to as `clusters', unless otherwise specified)
 arises when baryonic gas falls into
 the gravitational potential wells of hierarchically merging dark matter halos,
 and is shock heated to near the corresponding virial temperature.

Kaiser (1986) predicted that
 if the structure and evolutionary behavior of X-ray clusters
 are governed solely by gravitational processes,
 they should be `self-similar' and their physical properties
 should obey well-defined scaling laws.
Detailed numerical simulations (e.g. Bryan \& Norman 1998, Eke, Navarro \& Frenk 1998)
 also agree with this prediction.

However, observational evidence has been mounting
 that this is not the whole story.
Significant deviations from self-similarity are observed,
 most conspicuously in the X-ray luminosity-temperature (L-T) relation.
Whereas the self-similar model predicts
 the X-ray luminosity $L$ to scale with temperature $T$ as $L \propto T^2$,
 the observed relation is considerably steeper,
 being close to $L \propto T^3$ for $T \gtrsim $ 2 keV clusters
 (e.g. David et al. 1993, Allen \& Fabian 1998, Markevitch 1998, Arnaud \& Evrard 1999),
 and perhaps even steeper for lower $T$ groups
 (Ponman et al. 1996, Mulchaey \& Zabludoff 1998,
 Helsdon \& Ponman 2000, Xue \& Wu 2000).
Deviations have also been discovered in the entropy-temperature (S-T) relation,
 with evidence of an excess `entropy floor'
 in objects of $T \lesssim$ 2 keV
 (Ponman, Cannon \& Navarro 1999, Lloyd-Davies, Ponman \& Cannon 2000).

A plausible solution to these discrepancies
 is that the ICM
 (or the intergalactic medium prior to cluster formation
 which eventually becomes the present day ICM;
 also designated `ICM' for brevity)
 has been shock heated
 by some type of energy source
 in addition to gravitational heating
 (Kaiser 1991, Evrard \& Henry 1991;
 for alternative views, see Bryan 2000, Muanwong et al. 2001).
This process, while having little effect
 for the most massive and hottest clusters,
 can significantly modify lower $T$ systems,
 raising their temperature as well as decreasing their central gas densities
 from the self-similar expectations,
 and thereby bringing them into agreement with the observed scaling relations.
The most widely discussed energy source to date has been supernovae (SNe)
 (e.g. Wu, Fabian \& Nulsen 1998, 2000,
 Menci \& Cavaliere 2000, Cavaliere, Giacconi \& Menci 2000,
 Loewenstein 2000, Brighenti \& Mathews 2001).
As the ICM is observed to contain sub-solar abundances of Fe and other heavy elements,
 it is certain that metals produced by SNe inside cluster galaxies
 have been ejected into the ICM by some process,
 perhaps in the form of SN-driven `galactic winds'
 (e.g. Finoguenov, David \& Ponman 2000, Renzini 2000
 and references therein).
However, a number of recent studies have brought into question
 the effectiveness of SNe as {\em energy} sources for the ICM.
Employing semi-analytic models of cluster evolution,
 Valageas \& Silk 1999, Wu et al. 2000 and Bower et al. 2001
 have all shown that in order to reproduce the L-T relation,
 the non-gravitational heat input must amount to 0.5--3 keV per particle.
For supernovae,
 this implies an extremely high efficiency of energy conversion, close to 100\%,
 which is unrealistic considering that
 a large fraction of the initial energy is likely to be lost radiatively
 (e.g. Kravtsov \& Yepes 2000).
Some of the above papers have discussed the possibility
 that active galactic nuclei are instead the true heating agents of the ICM,
 but none have attempted a quantitative estimation of such an effect in any detail.

A related problem regards the soft X-ray background emission
 from the intergalactic medium (IGM),
 including regions outside virialized clusters.
Recent observational estimates (Fukugita, Hogan \& Peebles 1998)
 as well as numerical simulations
 (e.g. Cen \& Ostriker 1999, Dav\'e et al. 2001)
 suggest that the majority of the baryons in the universe today
 exists in a warm/hot IGM at temperatures $T \sim 10^5-10^7$ K. 
Concerns have been raised that the thermal emission from such regions
 may exceed the observed soft X-ray background
 in the absence of non-gravitational heating
 (Pen 1999, Wu, Fabian \& Nulsen 2001, Voit \& Bryan 2001, Bryan \& Voit 2001).
Again, the level of heating required to reconcile theory with observations
 may be of the order of a few keV per particle for the IGM.
However, this issue is currently controversial,
 as the inferences seem to depend critically on the numerical resolution and
 treatment of cooling in the calculations
 (Bryan \& Voit 2001 and references therein).

We consider in this work high-power, FRII radio galaxies (RGs)
 and shock heating of the ICM/IGM by their large-scale jets.
(The term `radio galaxy' is used here to connote all sources with strong jets
 including radio-loud quasars,
 which are believed to be intrinsically similar objects
 in the context of AGN unification schemes, Urry \& Padovani 1995.)
For Cyg A,
 the only nearby FRII RG residing inside a cluster
 that has been studied in sufficient depth,
 a strong case can be made
 that the RG is driving a bow shock into the ambient ICM
 (Carilli \& Barthel 1996 and references therein).
We wish to quantify the consequences of such effects for the ICM/IGM
 in a cosmological context.
As detailed observational information is available
 on the statistics of RGs in clusters (Ledlow \& Owen 1995, 1996)
 as well as on the evolution of RGs out to high redshifts
 (Dunlop \& Peacock 1990, Willott et al. 2001),
 we can obtain a reasonably reliable estimate
 of the total energy input into the ICM by RGs (Sec.\ref{sec:energy}).

We discuss the notable features and advantages of ICM heating by RGs
 (Sec.\ref{sec:rgheat}),
 as well as the validity of the assumptions
 and uncertainties that go into the above evaluation (Sec.\ref{sec:assum}).
A crucial diagnostic with which
 the RG scenario may be verified and distinguished from SN heating is proposed:
 the observation of high energy gamma-rays
 emitted by shock-accelerated non-thermal electrons
 during the period of shock activity (Sec.\ref{sec:gamma}).
Other implications,
 including non-thermal X-ray emission from groups,
 are also discussed (Sec.\ref{sec:other}).

A rough estimate of the RG energy input along the above lines
 have already been presented by Rawlings (2000).
En\ss lin et al. (1997, 1998)
 have pointed out the importance of RGs for the energetics of the ICM,
 and calculations similar to those given below
 were also carried out by En\ss lin \& Kaiser (2000);
 however, their focus was on the Sunyaev-Zeldovich effect,
 and they did not quantify the consequences
 in relation to the observed X-ray properties of clusters
 (cf. Yamada \& Fujita 2001).

\section{Energy Budget of Radio Galaxies}
 \label{sec:energy}

\subsection{Global Energy Output in the Universe}
 \label{sec:global}

We first evaluate the total energy output by RG jets in the universe,
 utilizing observational information
 on the evolution of the RG luminosity function (LF),
 and the correlation between jet kinetic power and radio luminosity of RGs.
Two cosmologies are considered,
 $\Omega = 1$, $\Lambda = 0$, $h_{50}=1$
 (Einstein-de Sitter, referred to as `EdS'),
 and
 $\Omega = 0.3$, $\Lambda = 0.7$, $h_{50}=1.4$
 (referred to as `lambda')
 where $H_0 = 50 h_{50} {\rm km \ s^{-1} Mpc^{-1}}$.
All radio luminosities are measured in units of ${\rm W \ Hz^{-1} sr^{-1}}$.

The redshift evolution of steep spectrum radio sources
 has recently been quantified in detail
 by Willott et al. (2001, hereafter W01).
Based on a large source sample with complete redshifts
 and less affected by various complicating effects
 due to the low frequency (151 MHz) selection,
 their data constitute the most accurate such information yet obtained.
Defined in terms of number of sources per unit comoving volume
 per unit logarithmic 151 MHZ radio luminosity $L_{151}$ at redshift $z$,
 their LFs $\rho(L_{151},z)$ were modeled as comprising two populations,
 a low-luminosity one and a high-luminosity one,
 corresponding to sources with weak and strong emission lines, respectively.
The break luminosity separating the populations,
 $\log L_{151} \sim 26.5$,
 is roughly one order of magnitude larger than the luminosity
 of the FRI/FRII division between the two morphological classes of RGs
 (Fanaroff \& Riley 1974).
The evolutionary behavior of each population was assumed to be different,
 with three different models being considered for the high-luminosity population:
 Model A (symmetric Gaussian in $z$),
 Model B (Gaussian rise up to a certain $z$ and constant beyond with no cutoff),
 Model C (Gaussian rise plus Gaussian decline in $z$ with different widths).
The low luminosity population was characterized as a power-law rise
 up to some $z$ and then constant beyond.
W01 fit such parameterizations to the data assuming two cosmologies,
 $\Omega =1$ or $\Omega =0$ with $\Lambda =0$, $h_{50}=1$,
 and found that all three models in either cosmology gave acceptable fits to the data
 including the `no cutoff' model B.
However, we will see that the total RG energy input
 is determined principally by RGs in the redshift range 
 well constrained by the data,
 and does not depend greatly on the uncertain high-$z$ evolution.
For our lambda cosmology, the LFs were obtained
 by converting the $\Omega =0$ LFs of W01 using their equation (14).

\begin{figure}
 \figurenum{1} 
 \epsscale{1.0} 
 \plotone{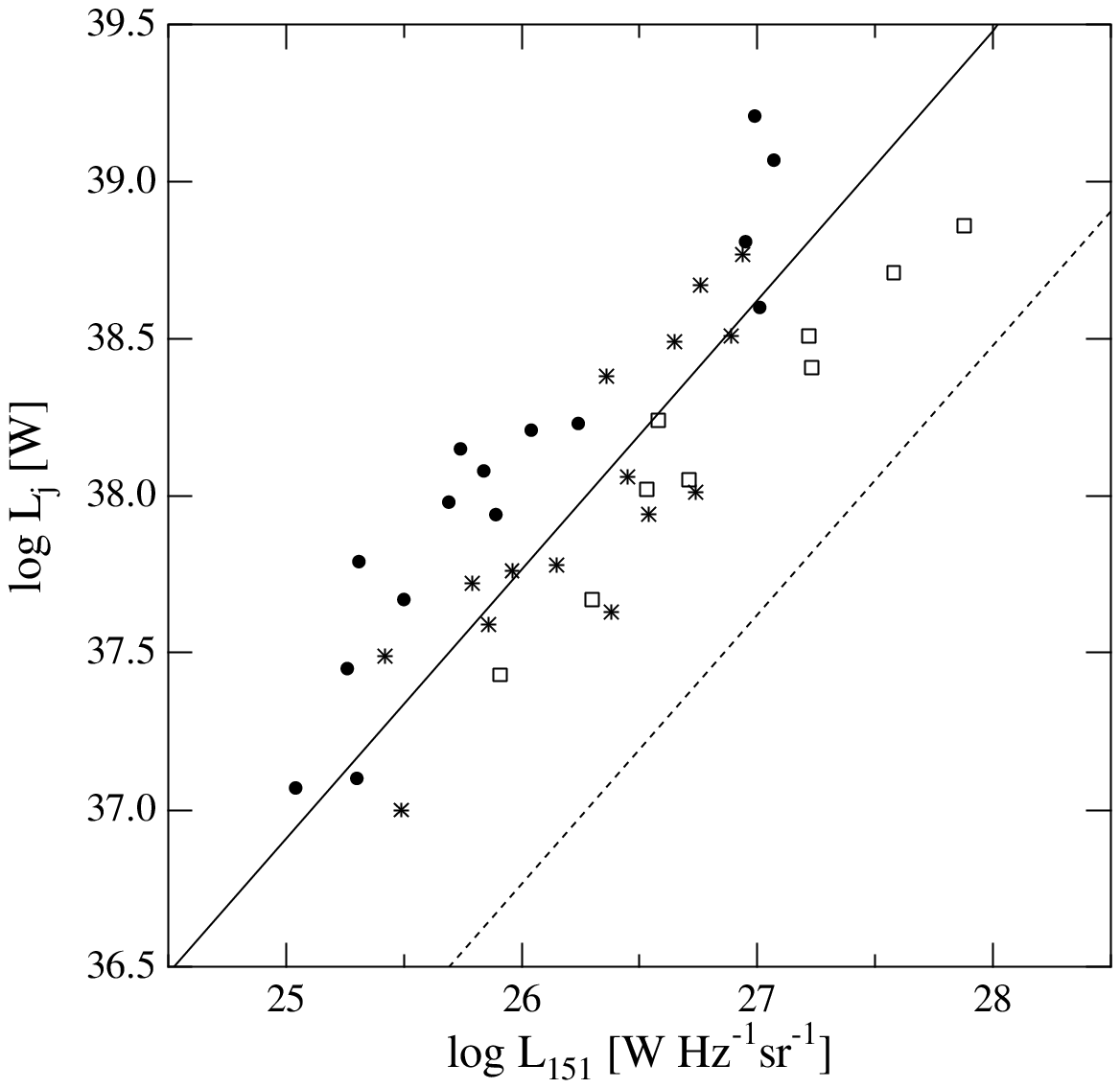} 
 \caption{
The observed correlation between $L_{151}$ and $L_j$ assuming $f_j=1$ and $h_{50}=1$,
 from Rawlings (1992).
RGs in isolated ($n \lesssim 10^{-5} \rm{cm^{-3}}$),
 group ($10^{-5} {\rm cm^{-3}} \lesssim n \lesssim 10^{-3} {\rm cm^{-3}}$),
 and cluster ($n \gtrsim 10^{-3} {\rm cm^{-3}}$)
 environments are indicated by circles, asterisks and squares, respectively.
Our adopted correlation of eq.\ref{eq:lj151} is indicated by the solid line,
 and that in Willott et al. (1999) by the dashed line.
} 
 \label{fig:lj}
 \end{figure}

Willott et al. (1999, hereafter W99),
 have derived a relation between the jet kinetic powers $L_j$ of RGs
 and their emitted radio luminosities at 151MHz $L_{151}$ of the form
 $L_j = 3 \times 10^{45} f_j (L_{151} / 10^{28} {\rm W \ Hz^{-1} sr^{-1}})^{6/7} {\rm erg \ s^{-1}}$.
This is based on physical considerations
 as well as on observational constraints of hot spot advance speeds and spectral ages
 for the general population of FRII RGs.
The important parameter $f_j$ takes into account
 uncertainties in the physical conditions inside the jet lobes
 (e.g. departures from equipartition, proton and low energy electron content,
 volume filling factor, etc.),
 and is expected to be in the range $f_j \sim 1-20$ (W99).
On the other hand, jet kinetic powers have also been determined observationally
 for individual RGs (modulo the uncertainty factor $f_j$) by Rawlings \& Saunders (1991),
 and the direct correlation between $L_j$ and $L_{151}$
 is available in Rawlings (1992) for RGs in different environments;
 this is shown in Fig.\ref{fig:lj}.
The W99 relation has the correct slope but
 seems to underpredict the magnitude of the observed $L_j$-$L_{151}$ relation.
We choose to adopt
 \begin{equation}
 L_j = 3 \times 10^{46}
       f_j \left({L_{151} \over 10^{28} {\rm W \ Hz^{-1} sr^{-1}} h_{50}^{-2}}\right)^{6/7}
       {\rm erg \ s^{-1}} h_{50}^{-2} ,
 \label{eq:lj151}
 \end{equation}
 agreeing well with the data for RGs in group-like environments
 (densities $n \sim 10^{-5}-10^{-3} {\rm cm^{-3}}$)
 that we are most interested in.
This modification to W99
 may imply that some of the parameters (e.g. the RG age)
 they discuss as being typical for powerful FRII RGs deviate from their actual values.
However, eq.\ref{eq:lj151}
 is not entirely inconsistent with the constraints discussed in W99,
 and is also in line with that given by En\ss lin et al. (1997)
 using radio luminosities at 2.7 GHz.
For $f_j$, we take a fiducial value of $f_j = 10$,
 which is supported by a number of independent observational inferences
 (e.g. Leahy \& Gizani 1999, Hardcastle \& Worrall 2000, Blundell \& Rawlings 2000;
 see Sec.\ref{sec:assum} for discussion of uncertainties in $f_j$).

Regarding individual RGs,
 $L_{151}$ should also depend on the RG's age even for constant $L_j$;
 however this dependence is expected to be weak (e.g. W99),
 allowing us to neglect this complication.
The density of the ambient medium obviously affects $L_{151}$ as well,
 but we assume that this does not induce any systematic effects,
 and that the above correlation is independent of $z$
 (see Sec.\ref{sec:assum} for justification).

\begin{figure}
 \figurenum{2} 
 \epsscale{1.0} 
 \plotone{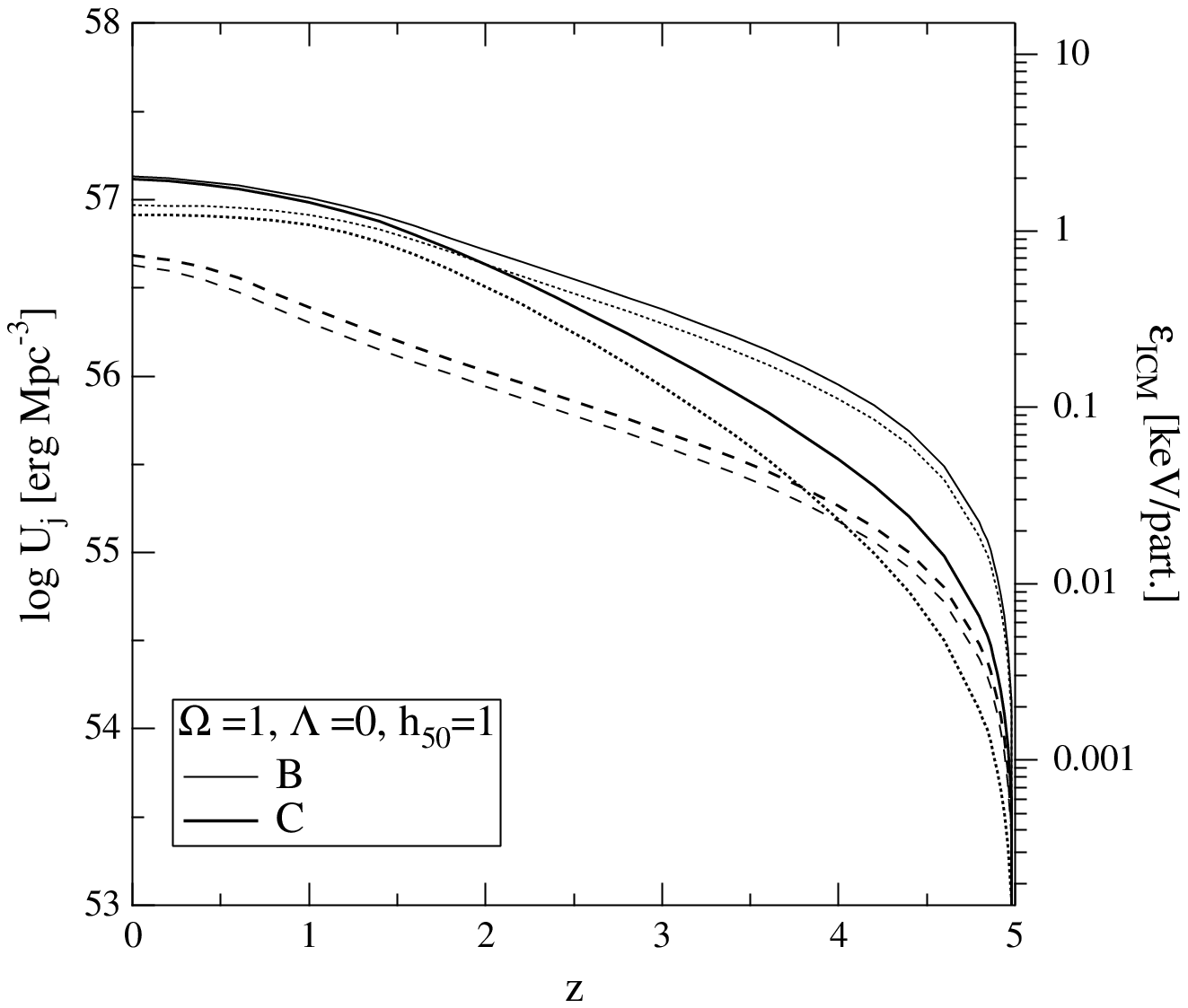} 
 \plotone{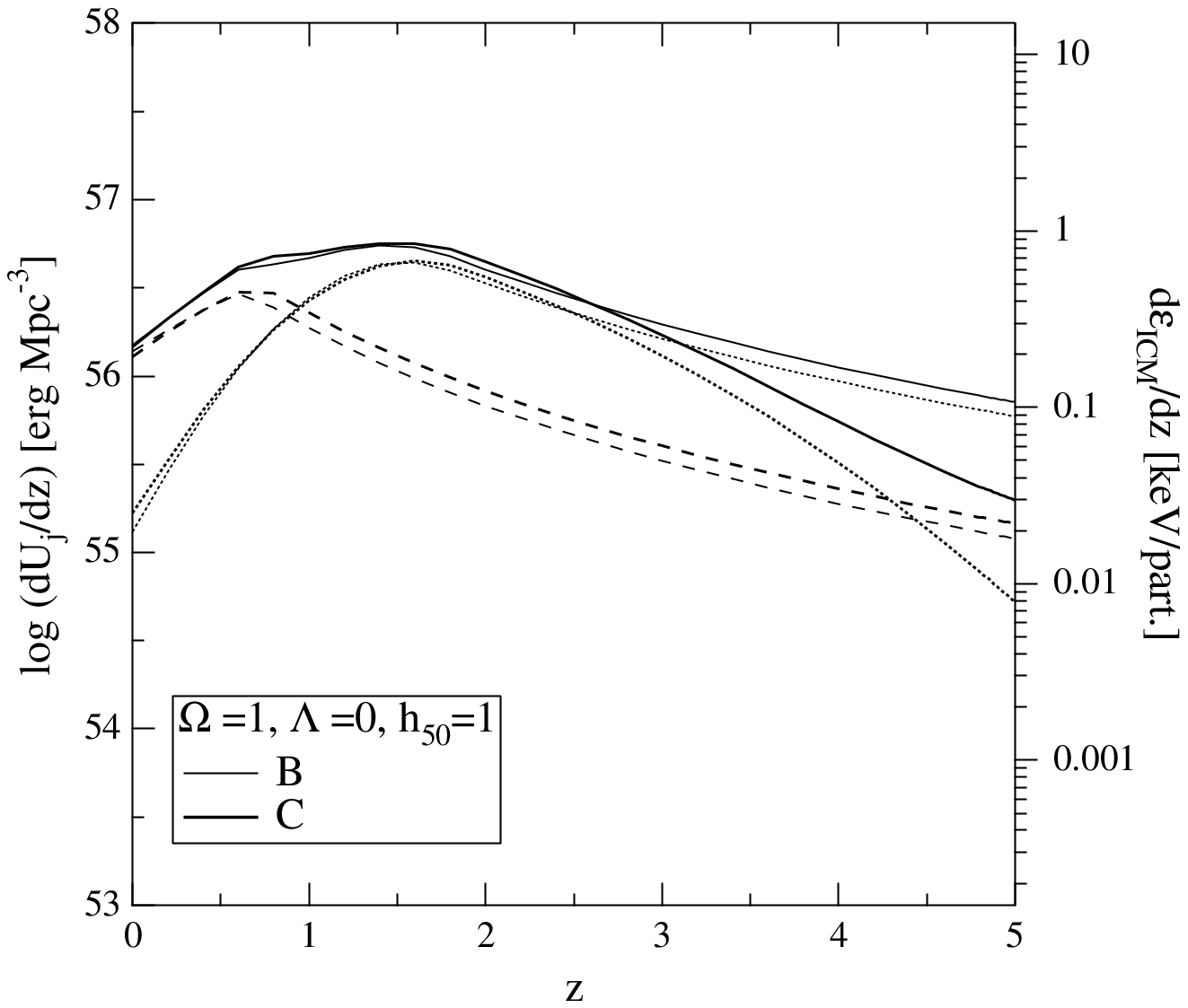} 
\caption{
a) The cumulative jet energy output from $z_{\max} =5$ up to $z$, and
b) The differential jet energy output per unit $z$,
 for model LFs B (thin) \& C (thick) of Willott et al. (2001)
 in the case of $\Omega=1$, $\Lambda=0$ and $h_{50}=1$.
The contributions of the low-luminosity population (dashed)
 and the high-luminosity population (dotted) are shown separately
 along with their sum (solid).
The right axes denote the corresponding energy input into the ICM.
}
 \label{fig:eds}
 \end{figure}

\begin{figure}
 \figurenum{3} 
 \epsscale{1.0} 
 \plotone{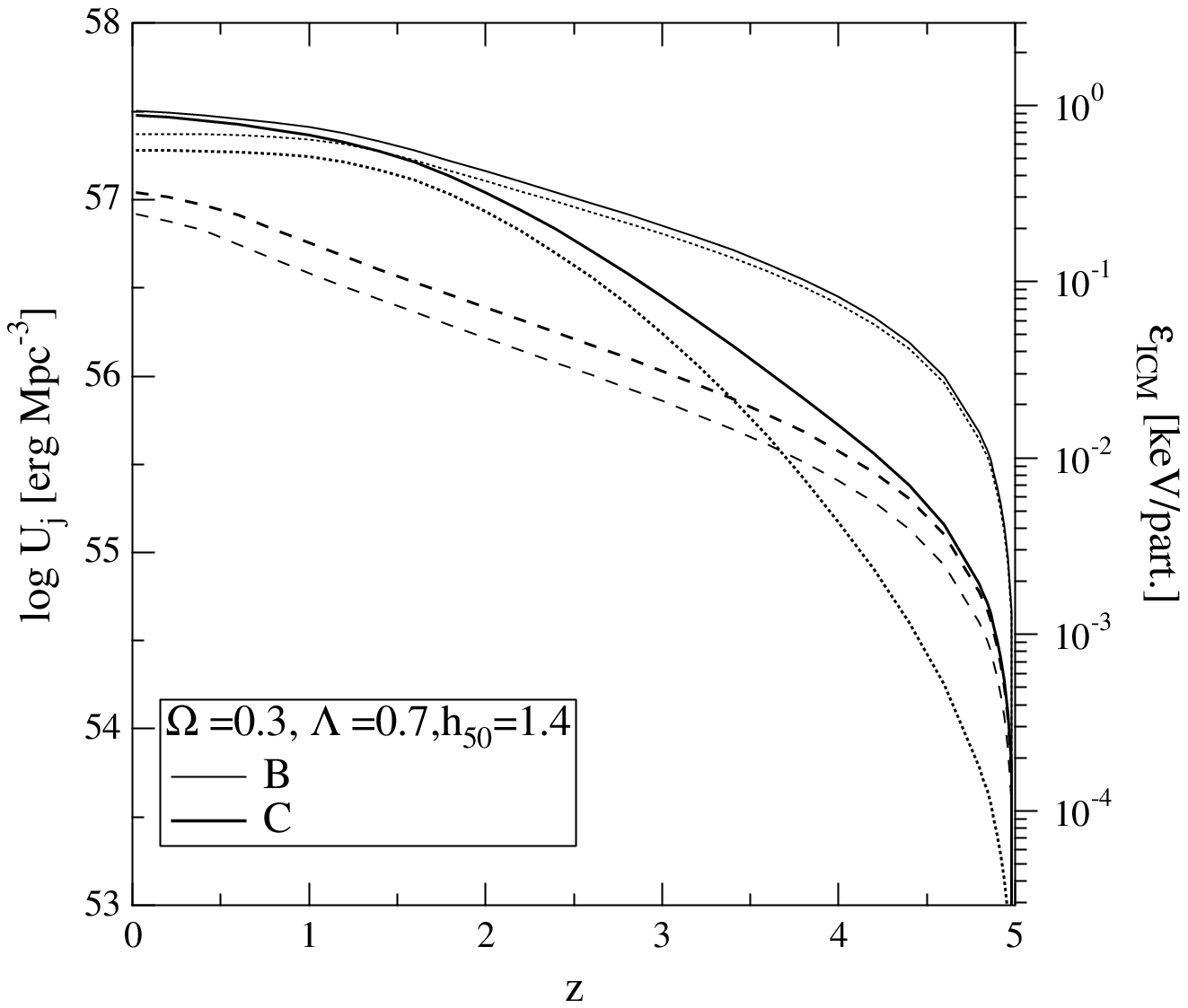} 
 \plotone{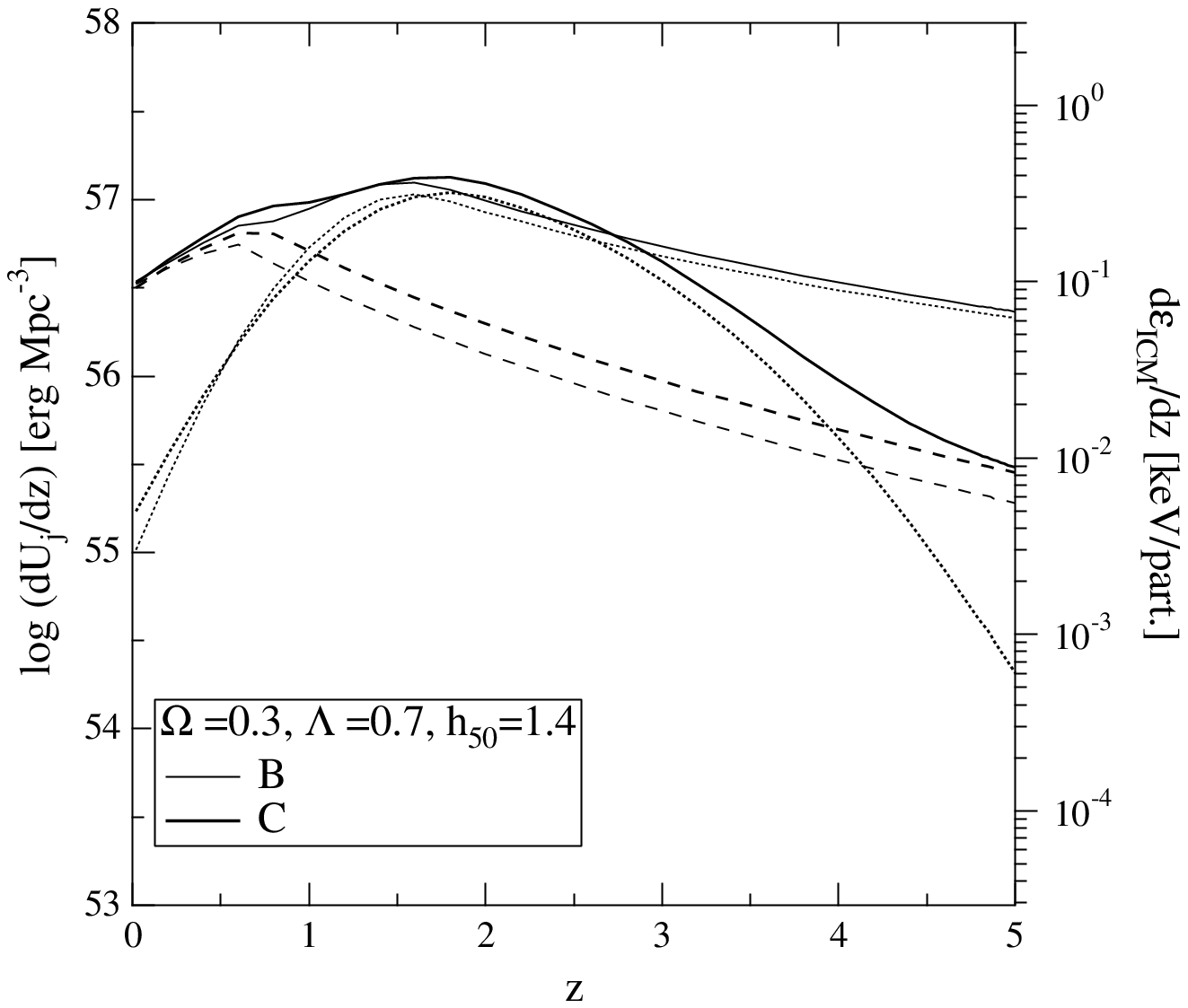} 
\caption{
Same as Fig.\ref{fig:eds}, but for $\Omega=0.3$, $\Lambda=0.7$ and $h_{50}=1.4$.
}
 \label{fig:lamb}
 \end{figure}

Utilizing the redshift dependent LFs and the above $L_j-L_{151}$ correlation,
 we integrate over radio luminosity and cosmic time
 to derive the total jet energy output by RGs $U_j$ in the universe,
 \begin{eqnarray}
 U_j = \int^{z_{\max}}_{z_{\min}} dz {dt \over dz}
              \int^{\log L_{151,\max}}_{\log L_{151,\min}} d\log L_{151}\\ \nonumber
              \times L_j(L_{151}) \rho(L_{151},z) .
 \label{eq:ljint}
 \end{eqnarray}
 (c.f. En\ss lin \& Kaiser 2000).
The lower and upper limits for $\log L_{151}$
 were taken to be 25.5 and 30, respectively.
The lower limit corresponds to the FRI/FRII dichotomy in radio morphology,
 below which RGs appear to be subsonic on large scales
 and do not drive strong shocks into the surrounding medium
 (e.g. Bicknell 1995, Laing 1996, Fabian 2001 and references therein).
The upper limit on $z$ was fixed to be $z_{\max}=5$,
 above which Compton cooling of RG lobes
 by the cosmic microwave background becomes very severe
 (Kaiser, Dennett-Thorpe \& Alexander 1997).
However, the results are not so sensitive to these values
 as $U_j$ is mainly determined by RGs
 near the `break' in the LF, $\log L_{151} \sim 26-27$,
 and in the redshift interval $z \sim 1-3$.
The results, both cumulative (integrated from $z_{\max}$ to $z$)
 and differential (per unit $z$),
 are shown in units of ${\rm erg \ Mpc^{-3}}$ in Figs.\ref{fig:eds} and \ref{fig:lamb}.

We see that the values of $U_j$ integrated to $z=0$
 amount to $1-3 \times 10^{57} {\rm erg \ Mpc^{-3}}$,
 and depend little on the LF model or on cosmology.
As can be seen from $dU_j/dz$ in Figs.\ref{fig:eds}b and \ref{fig:lamb}b,
 the dominant contribution to $U_j$ comes from sources at redshifts $z \sim$ 1--3,
 which are well constrained by the data
 and leave only small uncertainties in the integrals.
The numbers are also quite consistent with those obtained by En\ss lin \& Kaiser (2000)
 using different radio LFs and jet power-radio luminosity correlations.

We may average $U_j$ over the present-day baryon density in the universe
 and obtain
 \begin{equation}
 \epsilon_{\rm IGM} \simeq 0.15 {U_j \over 2 \times 10^{57} {\rm erg \ Mpc^{-3}}}
                               \left({\Omega_b h_{50}^2 \over 0.06}\right)^{-1} {\rm keV/particle} ,
 \label{eq:eigm}
 \end{equation}
 taking a mean molecular weight $\mu=0.6$.
If excess energies of a few keV/particle turn out to be necessary
 to halt overproduction of the soft X-ray background from the warm/hot IGM,
 (Sec.\ref{sec:intro}),
 RGs seem to fall short by at least an order of magnitude,
 even when assuming 100\% energy transfer from RG jets to the IGM.
However, more detailed calculations
 which account for the inhomogeneous IGM and RG clustering
 may be required to conclusively settle this matter.

If we consider groups in the mass range
 $M \sim 2 \times 10^{13}-2 \times 10^{14} M_\odot$
 (corresponding to $T \sim 0.5-2$ keV)
 in the context of currently favored structure formation models,
 their typical formation redshifts are
 $z_f \sim 0.3-0.5$ ($0.04-1.4$, 2 $\sigma$ dispersion range)
 for the EdS cosmology, and
 $z_f \sim 0.6-0.8$ ($0.1-1.9$, 2 $\sigma$ dispersion range)
 for the lambda cosmology
 (Lacey \& Cole 1993, Kitayama \& Suto 1996).
The main epoch of RG energy input, $z \sim 1-3$,
 therefore mostly precedes or coincides with the formation of low $T$ systems,
 and is generally consistent with the picture of `preheating',
 i.e. heating of the ICM prior to halo collapse.
Note also that this interval
 largely overlaps with the turnaround redshifts for these objects,
 at which the required amount of energy injection is minimal
 (e.g. Balogh, Babul \& Patton 1999, Tozzi \& Norman 2001);
 in comparison,
 the energy demands are much greater
 for heating after collapse and virialization
 (e.g. Tozzi, Scharf \& Norman 2000).

\subsection{Energy Input into the Intracluster Medium}
 \label{sec:icm}

We now translate the above results
 into the thermal energy input into the ICM gas,
 making use of the local LF of RGs inside clusters.
It is assumed that the redshift evolution of cluster RGs
 is similar to that of the whole RG population as determined by W01
 (see Sec.\ref{sec:assum}).
We begin by normalizing the local LF of W01
 to the LF determined for RGs inside clusters
 by Ledlow \& Owen (1995, 1996, hereafter respectively LO95, LO96),
 based on their 1.4 GHz VLA survey of Abell clusters.
Conversion between luminosities at 1.4 GHz and 151 MHz are done assuming
 a spectral energy index $\alpha =0.8$.
The LO96 univariate
 LF $f(L)$ is presented as the fraction of elliptical galaxies (excluding S0's)
 of R magnitude $R_c \le -20.5$
 detected at 1.4 GHz per logarithmic radio luminosity bin.
We determine the normalization factor ${\cal F}$ (units ${\rm Mpc^3}$)
 so that
 the W01 local LF $\rho(L,0)$ matches the LO96 LF,
 ${\cal F} \rho(L,0) \sim f(L)$,
 in the `break' luminosity range $\log L_{151} \sim $ 26--27.
(The quantity ${\cal F}^{-1}$ reflects the density of elliptical galaxies inside local clusters
 averaged over the universe,
 and ${\cal F} U_j$ corresponds to
 the average time-integrated jet energy input per cluster elliptical.)
We obtain $\log{\cal F} \sim 5.4$ for $h_{50}=1$ (EdS)
 and $\log{\cal F} \sim 4.5$ for $h_{50}=1.4$ (lambda).
The shape of the W01 local LF does not agree well with that of the LO96 LF
 for luminosities $\log L_{151} \lesssim 25$,
 but this is irrelevant for our purposes
 as our lower luminosity limit of integration is taken to be $\log L_{151} \sim 25.5$. 
We assume that this fractional LF at $z=0$
 is the same for all clusters and groups within the mass range of our interest
 (see Sec.\ref{sec:assum}).

It is further assumed that
 the number of elliptical galaxies $N_E$ in a cluster
 is proportional to its gas mass $M_g$.
More precisely,
 we posit that the gas which eventually evolves
 into the ICM of present-day clusters
 contains RG hosting ellipticals
 with a ratio of $N_E/M_g$ similar to that observed in local rich clusters.
This implies a mass-independent specific energy input,
 being the same for small groups as well as rich clusters. 
Taking the Coma cluster as a reference point,
 the number of elliptical galaxies of $R_c \le -20$
 within a radius $r \le 3.4 h_{50}^{-1} {\rm Mpc}$ from the cluster center
 is $\simeq 38$ (Thompson \& Gregory 1980).
The LO96 LF, when integrated over radio luminosity,
 gives a total fraction $\simeq 0.14$
 implying $\simeq 5$ RG hosting ellipticals in Coma;
 this compares favorably with
 the actual number of radio ellipticals observed with the appropriate properties
 in the Coma cluster (Venturi, Giovannini \& Feretti 1990).
The gas mass within the same radius is
 $M_g \simeq 4 \times 10^{14} h_{50}^{-1} M_\odot$ (Fusco-Femiano \& Hughes 1994),
 so we get a reference value of $N_E/M_g \simeq 0.95 \times 10^{-13} M_\odot^{-1}$.

The relativistic jet of an FRII RG should drive a strong shock
 into the ambient medium, heating and compressing it,
 but not all of the jet power can be directly conveyed outside in this manner.
A major portion of the total energy should also accumulate inside the `cocoon',
 the region immediately enveloping the jet lobes
 and filled with hot, shocked jet plasma,
 which is separated from the ICM by a contact discontinuity
 (Begelman \& Cioffi 1989).
Of the total energy $E_{\rm RG}$ released during a RG's lifetime,
 the fraction imparted to the surrounding ICM as work $E_{\rm ICM}$
 and that stored inside the cocoon as internal energy $E_c$
 can be evaluated in a simple way
 following En\ss lin \& Kaiser (2000, hereafter EK00).
After cessation of the RG activity, 
 the cocoon should expand up to a volume $V_c$ where its internal pressure $p_c$
 reaches equilibrium with that of the ICM $p_{\rm ICM}$, $p_c \sim p_{\rm ICM}$.
The work done against the ICM can be roughly estimated to be
 $E_{\rm ICM} \sim p_{\rm ICM} V_c$.
If the cocoon plasma is predominantly relativistic,
 $p_c \sim {1 \over 3} E_c/V_c$,
 and since $E_{\rm RG} \sim E_{\rm ICM} + E_c$,
 we get $E_{\rm ICM}/E_{\rm RG} \simeq 0.25$,
 whereas for primarily non-relativistic cocoon plasma,
 $p_c \sim {2 \over 3} E_c/V_c$ and $E_{\rm ICM}/E_{\rm RG} \simeq 0.4$.
As explained below (Sec.\ref{sec:rgheat}),
 the ICM gas shocked by the large-scale jets of FRII RGs
 should generally be hot and rarefied enough for
 radiative cooling to be negligible,
 and no further reduction in efficiency is expected.
We conservatively adopt $\xi_s=0.2$
 for the fraction of the total RG energy output imparted to the ICM.
(This value could be larger for reasons discussed in Sec.\ref{sec:assum}.)

Putting all of this together,
 the specific energy input into the ICM $\epsilon_{\rm ICM}$ is
 \begin{eqnarray}
 \epsilon_{\rm ICM} &=& \xi_s {\cal F} U_j \mu m_p N_E / M_g \\ \nonumber
          &\simeq& 1.2 {\xi_s \over 0.2} {{\cal F} \over 10^5 {\rm Mpc^3}}
                     {U_j \over 2 \times 10^{57} {\rm erg \ Mpc^{-3}}}\\ \nonumber
          &\times& {N_E/M_g \over 0.95 \times 10^{-13} M_\odot^{-1}} \ \ {\rm keV/particle} ,
 \end{eqnarray}
 using $\mu=0.6$.
In Figs.\ref{fig:eds}a and \ref{fig:lamb}a,
 the right axes denote the values of $\epsilon_{\rm ICM}$
 calculated for the different LF models.
We see that with reasonable assumptions,
 the total energy input into the ICM by RGs lies in the range
 $\sim$ 1--2 keV per particle
 (and could be even higher, see Sec.\ref{sec:assum}).
According to recent detailed models of cluster evolution
 incorporating non-gravitational feedback
 (Valageas \& Silk 1999, Wu et al. 1998, 2000, Bower et al. 2001,
 Brighenti \& Mathews 2001),
 a few keV per particle is just the amount of excess energy input required
 to account for the observed deviations from self-similarity in X-ray clusters.

\section{Discussion}
\label{sec:disc}

\subsection{On Heating of the Intracluster Medium by Radio Galaxies}
\label{sec:rgheat}

We discuss some important issues regarding
 heating of the ICM by RGs.
First,
 a single FRII RG is capable of sweeping out and shock heating a volume
 of order $({\rm Mpc})^3$ during its lifetime.
This is enough to affect a major fraction of the total ICM volume,
 especially for relatively low $T$ clusters and groups.
We also mention that RGs
 are observed to possess a more centrally concentrated
 spatial distribution relative to normal galaxies,
 at least for relatively rich clusters (LO95).

Next, we emphasize the crucial advantage of RGs compared to SNe for ICM heating.
The jets of high-power, FRII RGs are known to transport their energy efficiently
 out to regions well outside their host galaxies,
 and then deposit it directly into the low-density, large-scale ICM.
The shocked gas is expected to possess
 a sufficiently low density and high temperature
 such that its radiative cooling time is much longer than the Hubble time
 (Heinz, Reynolds \& Begelman 1998, Kaiser \& Alexander 1999),
 possibly apart from the innermost regions of rich clusters
 constituting a minor fraction of the whole ICM (Tozzi \& Norman 2001).
This is in stark contrast to SNe,
 which must first explode into the dense interstellar medium of their host galaxies
 where the cooling time is much shorter (see however Sasaki 2001),
 and thereby lose a major fraction of their energy as escaping radiation
 (Kravtsov \& Yepes 2000, Wu et al. 2000).

In relation to the above two points,
 we note that the lifetimes of individual FRII RGs
 are believed to be $\lesssim 10^8$ yr (e.g. Blundell, Rawlings \& Willott 1999),
 much shorter than the main duration of activity for the FRII population as a whole,
 as well as the typical ages of clusters.
By virtue of the long cooling time of the shocked ICM,
 any FRII activity in a cluster's past will have left its mark on the ICM,
 but need not have any direct correspondence
 with the degree of present-day RG activity in a particular cluster.
It is also to be noted that for small groups,
 which typically contain only a few large elliptical members,
 the LO96 fractional local LF nominally implies RG numbers less than one.
This is to be interpreted as the present-day probability of observing 
 RG activity in these systems;
 e.g. for a mass range in which 0.1 RGs are implied,
 one should see only 1 out of 10 groups to harbor a RG.
However, the occurrence of FRII RGs increases dramatically at high $z$,
 and we expect that all sufficiently large ellipticals
 have hosted an FRII RG at some point in their lives (c.f. LO96).

We also note that X-ray observations
 of a number of presently active RGs inside nearby clusters,
 including recent high-resolution data by {\it Chandra},
 do not show any evidence of the RG heating the external medium
 (Fabian 2001 and references therein).
However, one must remember that nearly all such nearby objects,
 including M87 in Virgo or 3C84 in Perseus,
 are relatively low-power, FRI-type RGs,
 which are presumably subsonic on large scales
 and incapable of shock heating the ambient ICM (e.g. Bicknell 1995, Laing 1996);
 such RGs were not included in our evaluation above.
FRII RGs in dense environments
 are known to be quite rare in the local universe,
 with the one exception, Cyg A,
 displaying good evidence for a bow shock being driven into the ICM,
 compressing and heating it
 (Carilli, Perley \& Dreher 1988, Carilli, Perley \& Harris 1994,
 Clarke, Harris \& Carilli 1997, Wilson, Young \& Shopbell 2001b).
The prominence of FRII RGs at high redshifts
 does not entail any contradictions
 in the role of RGs in shock heating the ICM 
 with observations of nearby cluster RGs.

It is to be mentioned that
 besides the process of shock heating,
 which endows the ICM with excess thermal energy and entropy,
 a `gravitational' form of the surplus energy may arise
 if RGs can displace the ICM gas sufficiently outward in the cluster potential
 at appropriate epochs before cluster virialization
 (see Sec.7.1 of Wu et al. 2000).
For low mass clusters,
 RGs may also entirely unbind parts of the ICM from the potential,
 reducing the gas fraction and helping to steepen the L-T relation.
Note that since this spatial redistribution can be realized adiabatically without shocking,
 even FRI RGs become potential contributors to self-similarity breaking.
However, such processes alone may be inadequate
 to account for the observed magnitude of the entropy floor in the S-T relation,
 so that shock heating by FRII RGs is still essential.
(FRII RGs should also be more efficient at expelling the ICM gas anyhow.)
A quantitative assessment of these effects requires more detailed modeling,
 as well as better observational data on the scaling relations.

\subsection{On Assumptions and Uncertainties}
\label{sec:assum}

We have obtained numbers for the excess energy input
 of $\epsilon_{\rm ICM} \sim$ 1 keV/particle
 using what we deem to be the most plausible assumptions
 and reasonable parameter values.
Some remarks concerning these are in order.

We first call attention to three key assumptions
 used in the above evaluation:
 1) the $L_j-L_{151}$ correlation is independent of $z$;
 2) the redshift evolution of RGs in clusters
 is the same as that for the whole RG population; and
 3) the fractional RG LF is the same for all clusters.
Assumption 1) is consistent with inferences
 drawn by Blundell et al. (1999),
 who find no systematic epoch-dependence
 in the gaseous environments of the RG population as a whole.
Recent investigations of the clustering environments of RGs
 (e.g. Wold et al. 2000, McLure \& Dunlop 2001)
 do not point to a strong influence of the environment
 on the evolutionary properties of RGs,
 compatible with assumptions 1) and 2).
For assumption 3),
 the Ledlow-Owen fractional LF is independent of cluster richness or morphological type
 within in their sample of Abell clusters (LO95),
 and it is not unreasonable to expect
 that this extends down to the scale of small groups.
These are certainly no proofs, but lend some credence to our assumptions.

Next we consider the uncertainties in some of our parameters
 that directly affect $\epsilon_{\rm ICM}$.
Although the fiducial value of $f_j=10$ was chosen to conform with
 observational inferences made by several authors using different methods
 (Leahy \& Gizani 1999, Hardcastle \& Worrall 2000, Blundell \& Rawlings 2000),
 it is warned that the associated uncertainty could be rather large,
 perhaps by a factor of a few or more.
We note that detections of synchrotron-self-Compton X-rays
 from the hot spots of some FRII RGs suggest that in these objects
 the magnetic fields are close to equipartition with the radiating electrons
 (Harris, Carilli \& Perley 1994, Wilson, Young \& Shopbell 2000,
 Harris et al. 2000, Hardcastle, Birkinshaw \& Worrall 2001).
At face value, this may indicate $f_j \simeq 1$,
 but some caveats are to be mentioned:
 i) even if the hot spot is in equipartition,
 significant deviations from it may arise inside the whole radio lobe,
 as discussed by Blundell \& Rawlings (2000)
 who argue for $f_j \simeq 10$ in FRII RGs;
 ii) there are no convincing reasons a priori that equipartition must hold,
 so the possibility is open that
 the hot spots and lobes carry a substantial amount of energy in non-radiating particles
 (such as protons or low-energy electrons)
 by violating equipartition (Hardcastle \& Worrall 2000).
Indeed, measurements of the thermal pressure of the external medium of RGs,
 arguably being a more direct probe of the total lobe energy content,
 show that in general FRII RG lobes would be highly underpressured
 with respect to the ambient gas if equipartition prevails
 (Clarke, Harris \& Carilli 1997, Leahy \& Gizani 1999, Hardcastle \& Worrall 2000).
This would be at odds with our current understanding of FRII RGs,
 which requires the lobe pressures to be at least as high as the external medium
 (e.g. Begelman \& Cioffi 1989, Kaiser \& Alexander 1997 and references therein),
 so that i) and/or ii) of the above may be occurring in reality.
We also add that there are some FRII RGs whose hot spot X-ray emission 
 cannot be readily explained with equipartition parameters
 (e.g. Wilson, Young \& Shopbell 2001a).
Further detailed observations
 of RG hot spots and lobes and of the external confining medium of RGs
 should help to nail down the value of the crucial parameter $f_j$.

We also need to
 check that the ratio $N_E/M_g$ of the Coma cluster we have used
 is representative of the global value,
 as well as to substantiate the assumption
 that this is relatively constant within the cluster mass range of interest.
Here we simply remark that this number could also be rather uncertain,
 and warrants better constraints from future observations with improved statistics.

In contrast to parameters deduced from observations,
 $\xi_s$ was evaluated theoretically,
 and we point to one uncertain aspect which can increase this value.
Our discussion above was limited
 to heating of the ICM by the forward shocks induced by RGs,
 but as mentioned in Sec.\ref{sec:icm},
 the better part of the total RG energy release
 should be stored inside its cocoon as internal energy of jet plasma.
Although it is believed that the the cocoons of FRII RGs
 are overpressured and their interior segregated from the ICM
 during much of the RGs' lifetime (Begelman \& Cioffi 1989),
 it is unclear what occurs after the activity ceases.
One possibility is that cocoons mostly dissipate
 after reaching pressure equilibrium with the ambient ICM
 due to instabilities operating at the cocoon-ICM interface
 (or perhaps even earlier, Reynolds, Heinz \& Begelman 2001).
Alternatively, it has been proposed that
 cocoons may retain their entity for a long time afterwards,
 perhaps until the host cluster undergoes a major merger (EK00).
If the cocoon dissipates,
 its internal energy, amounting to 75\% (60\%)
 of the total RG energy release for relativistic (non-relativistic) cocoon plasma,
 can be released into the ICM,
 and any fraction that is thermalized with the ICM
 will add to the excess energy and raise $\xi_s$.
The actual amount that can contribute to $\epsilon_{\rm ICM}$ in this way
 is difficult to assess,
 as it depends on the highly uncertain composition, particle distribution,
 and magnetic field content of the jet plasma,
 but in any case, we should bear in mind
 the fact that the RG cocoon is a significant energy reservoir.

\subsection{Comments on Heating by Supernovae}
\label{sec:snheat}

As discussed in Sec.\ref{sec:intro},
 the substantial amounts of heavy elements observed in the ICM
 certifies that SNe must have affected the ICM in one way or other,
 but this does not immediately imply that
 they have been energetically important.
One may envision ways of polluting the ICM with metals
 without recourse to SN-driven galactic winds,
 such as ram-pressure stripping,
 repeated tidal interactions (`galaxy harassment'), mergers, etc.
Theoretical models, at least in their simplest forms,
 face difficulty in transporting
 the combined kinetic energy of an ensemble of SNe
 out from their host galaxies and into the ICM with high efficiency;
 severe radiative losses are to be expected during this process
 (Thornton et al. 1998, Kravtsov \& Yepes 2000).

However, observations of
 SN-driven winds in starburst galaxies
 indicate a relatively high efficiency of energy conversion (Heckman 2000),
 and theoretical consideration of a realistic, multiphase nature
 of the starburst region into which SNe explode 
 may alleviate adverse radiative energy loss (Strickland 2001).
Thus the view
 that SN-driven winds play a significant role in the energetics of the ICM
 cannot be ruled out.
Sasaki (2001) has also suggested the possibility of intracluster SNe,
 i.e. those which explode outside galaxies and directly heat the ICM.
It is then preferable to have some independent observational test
 which can discriminate between our proposal of RGs heating the ICM
 and that by SNe.

In principle,
 one method involves the redshift evolution of the X-ray scaling relations.
The space density of FRII RGs
 is seen to increase with redshift faster than the average star formation rate
 (approximately equal to the SN rate) in the universe
 (e.g. Madau 1999 and references therein),
 and this may be reflected in the $z$ evolution of the L-T and S-T relations.
At present the L-T relation is available only out to $z \sim 0.4$
 and only with large errors (Mushotzky \& Scharf 1997),
 but the observational situation should improve significantly in the near future
 with {\it Chandra} and {\it XMM-Newton} observations.
However, the expected difference between RGs and SNe may not be too drastic,
 and could render this prospect impractical.

One would appreciate some observational signature
 which can directly identify RGs in the process
 of heating the ICM at the expected epochs.
The long radiative cooling time of the shocked ICM
 means that the thermal X-ray emission from this gas
 never becomes very luminous
 and could be difficult to observe.
However, the same shocks which heat the ICM should
 also generate non-thermal particles through shock acceleration,
 and the associated high energy radiation,
 particularly from relativistic electrons,
 can offer a potent diagnostic,
 which we discuss next.

\subsection{Non-thermal Gamma-ray Emission}
\label{sec:gamma}

Non-thermal gamma-rays
 emitted by shock accelerated electrons
 hold the promise of being a very direct and powerful probe
 of large-scale shock heating processes in the ICM/IGM.
Strong shocks, which are a requisite element in our picture of ICM heating,
 are also known to be very conducive to non-thermal particle acceleration
 via the first order Fermi mechanism
 (Blandford \& Eichler 1987 and references therein).
Relativistic electrons accelerated and injected into the shocked ICM
 subsequently lose energy by emitting synchrotron and inverse Compton (IC) radiation,
 the latter mainly by upscattering cosmic microwave background photons.
For typical ICM magnetic fields ($B_{\rm ICM} \sim 0.1-1 {\rm \mu G}$),
 IC dominates, and much of the electrons' energy
 should end up as gamma-rays.
The crucial point of note is that
 at sufficiently high energies, relativistic electrons
 have IC cooling times much shorter than the duration of the shock (here the RG lifetime),
 so that the resultant gamma-ray emission is `prompt'
 and directly traces the period of strong shock activity.
Such gamma-rays therefore serve as beacons,
 enabling us to `see' the moment
 at which large-scale shocks are occurring.
The significance of such prompt gamma-ray emission
 was recognized by Loeb \& Waxman (2000) and Totani \& Kitayama (2000),
 in the context of gravitational shocks in groups and clusters.
Gamma-rays of gravitational shock origin may be suppressed
 if, as in our RG scenario, the pre-collapse IGM of groups and clusters
 has been substantially heated by non-gravitational sources (Totani \& Inoue 2001).

For shocks induced by the RGs themselves,
 the typical gamma-ray spectra and luminosities can be estimated as follows.
A fraction $\xi_e$ of the RG kinetic power $L_j$
 is assumed to be converted to non-thermal electrons
 with a power-law distribution $dn/d\gamma \propto \gamma^{-p}$,
 $\gamma$ being the electron Lorentz factor.
The maximum Lorentz factor $\gamma_{\max}$ is set
 by equating the shock acceleration time with the IC cooling time,
 $\gamma_{\max} \simeq 1.2 \times 10^9 (1+z)^{-2}
                     (B_{\rm ICM} / 1 {\rm \mu G})^{1/2} (V_s / 10^4 {\rm km s^{-1}})$,
 where $V_s \simeq 10^4 {\rm km s^{-1}}$
 is a typical velocity of the forward shock ($\simeq$ the hot spot advance speed)
 for an FRII RG.
The cooling Lorentz factor $\gamma_c$ is where
 the IC cooling time equals the RG lifetime $\tau_{\rm RG} \sim 10^8 {\rm yr}$,
 $\gamma_c \simeq 2.3 \times 10^4 (1+z)^{-4} (\tau_{\rm RG} / 10^8 {\rm yr})^{-1}$,
 and above this Lorentz factor all electrons cool within the shock duration.
The emission energies corresponding to $\gamma_c$ and $\gamma_{\max}$
 are respectively
 $E_c \simeq 450 {\rm keV} (1+z)^{-7} (\tau_{\rm RG} / 10^8 {\rm yr})^{-2}$ and
 $E_{\max} \simeq 1.2 \times 10^3 {\rm TeV} (1+z)^{-3}
                (B_{\rm ICM} / 1 {\rm \mu G}) (V_s / 10^4 {\rm km s^{-1}})^2$,
 between which the spectrum will have a cooled energy index $\alpha_c = p/2$.
Below $\gamma_c$ the electrons are mostly adiabatic,
 and the energy index of the spectrum below $E_c$ is $\alpha_{ad} = (p-1)/2$.

Strong shocks should result in $p \simeq 2$
 and hence $\alpha_{ad} \simeq 0.5$ and $\alpha_c \simeq 1$,
 i.e. constant luminosity per logarithmic energy interval above $E_c$.
The efficiency of electron injection $\xi_e$ in astrophysical shocks
 is a critical but rather uncertain parameter,
 both observationally and theoretically.
For shocks in supernova remnants,
 this is believed to be somewhere in the range $\xi_e \sim 0.001-0.05$
 (e.g. Sturner et al. 1997, Baring et al. 1999, Tanimori et al. 1998).
It may also be more appropriate to measure $\xi_e$
 with respect to the thermal energy of the post-shock gas
 rather the total kinetic energy,
 in which case we must also account for $\xi_s$,
 the efficiency of energy input into the ICM.
Although our fiducial choice will be $\xi_e = 0.05$,
 we caution that this number could be lower in reality,
 and our estimates below
 rather optimistic.
We do note that
 recent observations of hard X-ray tails in groups and clusters (e.g. Rephaeli 2001),
 if interpreted as non-thermal emission from electrons,
 seem to require high electron injection efficiencies, of order $\xi_e = 0.05$
 (e.g. Sarazin 1999, Takizawa \& Naito 2000, Blasi 2001).

A RG at $z$ will be observed
 to have an integral gamma-ray flux above energy $E_*$
 \begin{equation}
 F_* \sim {1+z \over E_* \ln(\gamma_{\max})} {\xi_e L_j \over 4\pi d_L(z)^2} ,
 \end{equation}
 where $d_L(z)$ is the luminosity distance (c.f. Totani \& Kitayama 2000).
The nearest, luminous FRII RG Cyg A
 lies in a relatively rich cluster at $z=0.058$,
 and should possess a jet power $L_j \simeq 3 \times 10^{46} {\rm erg s^{-1}}$.
In the lambda cosmology,
 the high-energy gamma-ray flux above 100 MeV from this object
 is estimated to be $F_{100} \simeq 8 \times 10^{-8}$ ${\rm photons \ cm^{-2} s^{-1}}$,
 below the 2 $\sigma$ upper limit
 ascertained by the {\it EGRET} instrument,
 $1.7 \times 10^{-7}$ ${\rm photons \ cm^{-2} s^{-1}}$ (Fichtel et al. 1994).
However,
 this is easily detectable by the {\it GLAST} satellite,
 whose nominal sensitivity for a 2 year all-sky survey
 is $2 \times 10^{-9}$ ${\rm photons \ cm^{-2} s^{-1}}$ above 100 MeV
 (Gehrels \& Michelson 1999);
 other relatively nearby FRII RGs should be observable as well.
The more luminous high-redshift RGs
 with radio luminosities $L_{151} \gtrsim 10^{28} {\rm W \ Hz^{-1} sr^{-1}}$
 and corresponding jet powers $L_j \gtrsim 2.7 \times 10^{47} {\rm erg s^{-1}}$
 would be within the capabilities of {\it GLAST} even at $z=1$,
 allowing the era of substantial ICM heating to be directly probed.

\begin{figure}
 \figurenum{4} 
 \epsscale{1.0} 
 \plotone{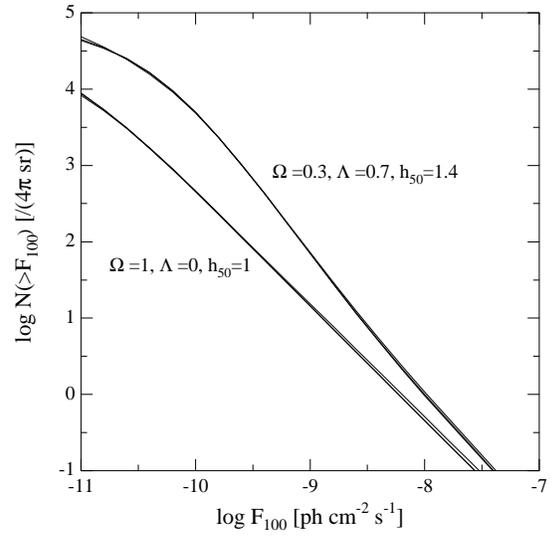} 
 \caption{
The cumulative all-sky number of RGs with gamma-ray flux larger than $F_{100}$.
Curves are shown for all model LFs (A, B and C),
 but are almost degenerate within either cosmology, EdS (bottom) and lambda (top).
} 
 \label{fig:gamct}
 \end{figure}

Using the W01 radio LFs,
 it is also straightforward to evaluate the gamma-ray source counts
 for the whole population.
The number of RGs with gamma-ray flux greater than $F_*$ is
 \begin{eqnarray}
 N(>F_*) &=& \int dz {dV \over dz}
             \int^{\log L_{151,\max}}_{\max\{\log L_{151}(F_*,z), \log L_{151,\min}\}}\\ \nonumber
         &\times& d\log L_{151} \rho(L_{151},z) ,
 \end{eqnarray}
 where $L_{151}(F_*,z)$ is the 151 MHz radio luminosity of a RG
 at $z$ that would produce a flux $F_*$,
 and ${dV \over dz}$ is the comoving volume element of the universe.
Fig.\ref{fig:gamct} shows the predicted $\log N-\log F_{100}$ distribution
 of gamma-ray emitting RGs.
For the lambda cosmology,
 we predict that {\it GLAST} would see
 about 20 discrete gamma-ray sources
 which are non-variable and associated with powerful RGs
 during a 2 year all-sky survey.
Since these RGs are strong radio emitters as well,
 targeted observations of selected objects could do even better;
 for example, at a sensitivity of $10^{-9}$ ${\rm photons \ cm^{-2} s^{-1}}$
 (just a factor of 2 better than the 2 year survey limit),
 up to 70 RGs are detectable.

On the other hand,
 the number of observable RGs should have been negligible for {\it EGRET}.
This indicates that RGs are unlikely to have contributed greatly
 to any of the unidentified EGRET sources (Hartman et al. 1999)
 or the extragalactic gamma-ray background (Sreekumar et al. 1998).

It is emphasized that
 besides the forward shock in the ICM,
 electron acceleration should also be conspicuous
 at the reverse shock inside the radio lobes, i.e. the hot spots.
However, copious high energy gamma-ray emission is expected only from the forward shock;
 since the magnetic fields in the hot spots are much higher than in the ICM
 (e.g. Meisenheimer et al. 1989),
 synchrotron cooling dominates
 and forces the maximum emission energy of the reverse shock electrons
 to be well below the {\it GLAST} energy band.
This makes the above test all the more effective:
 high energy gamma-ray emission
 does not merely signal the presence of a powerful RG {\it per se},
 but necessitates that the external medium is being actively shocked.

The gamma-ray diagnosis should be markedly different
 if ICM heating is due to SN-driven winds.
The kinetic powers of individual winds
 is probably at most $L_w \sim 10^{42}-10^{43} {\rm erg \ s^{-1}}$,
 and their lifetimes may be around $\tau_w \sim 10^7 {\rm yr}$
 (Heckman 2000).
Thus the consequent non-thermal emission
 by shock-accelerated electrons of SN-driven wind origin
 is much too weak to be detectable as discrete gamma-ray sources,
 barring the highly unlikely situation of $10^3$ to $10^4$ such winds
 being active nearly simultaneously (i.e. within $\tau_w$).
These statements apply to intracluster SNe as well,
 whose explosion times should be even less correlated with each other.
Thus future gamma-ray observations offer good prospects
 for directly corroborating that
 RGs indeed shock heat the ICM at the expected epochs,
 as well as discriminating between the SNe scenario.

We further remark that the same shocks which accelerate electrons 
 should do so for protons as well,
 perhaps with an even higher injection efficiency.
These non-thermal protons can undergo
 inelastic collisions with thermal protons of the ICM gas,
 and thereby emit high energy gamma-rays via neutral pion decay
 (e.g. Dar \& Shaviv 1995, V\"olk, Aharonian \& Breitschwerdt 1996,
 Berezinsky, Blasi \& Ptuskin 1997),
 or through IC radiation by secondary electrons
 (e.g. Blasi \& Colafrancesco 1999).
As the cooling times of such high energy protons
 are generally longer than the Hubble time (En\ss lin et al. 1997),
 proton-induced gamma-ray emission is neither efficient nor prompt,
 and would not serve as useful `signposts' of the shock heating epoch.
On the other hand,
 their confinement times in the ICM are also expected to be very long
 (V\"olk et al. 1996, Berezinsky et al. 1997),
 so any non-thermal protons injected into the ICM by various kinds of shock activity
 may accumulate over the lifetime of a cluster,
 and the accompanying emission should carry
 important information on its entropy history.
We wish to explore this interesting issue
 of non-thermal protons in relation to
 non-gravitational mechanical feedback processes in the IGM/ICM
 in a future paper.

\subsection{Non-thermal X-ray Emission and Other Implications}
\label{sec:other}

We briefly remark on some additional implications.

Following the above discussion, we note that
 relativistic electrons accelerated at the ICM shock
 should also emit non-thermal X-ray emission.
These X-rays will not be as effective as gamma-rays
 in offering a test of our proposal,
 since 1) the emitting electrons have cooling times longer than the shock duration,
 so most of the emission is not `prompt'
 and delayed with respect to the RG activity,
 and 2) confusion with other X-ray components,
 such as emission from the RG hotspots and knots, the nucleus,
 as well as from the undisturbed ICM, is problematic.
Conversely,
 non-thermal X-ray emission is interesting in that
 it may last for a relatively long time
 after the shock activity and particle injection has ceased,
 and the implications of such `relic' X-ray emission
 has been discussed in several recent papers
 (e.g. Sarazin 1999, Atoyan \& V\"olk 2000, Blasi 2001)
 in response to possible detections of hard X-ray tails in rich clusters
 (Rephaeli 2001 and references therein).
With regard to our study,
 we call attention to recent reports of excess hard X-ray emission
 in the nearby group of galaxies HCG62 by Fukazawa et al. (2001),
 and similar detections for other groups by Nakazawa (2001, private communication).
In any non-gravitational heating scenario
 which envisions significant shock heating of group gas prior to its formation,
 the gravitational shock accompanying the subsequent collapse
 is strongly suppressed (e.g. Balogh et al. 1999, Tozzi \& Norman 2001).
Non-thermal particle acceleration should result
 only from the heating sources themselves, which are RGs in our case.
The radio emission from the RG itself
 could fade rather rapidly after its death
 (Goldshmidt \& Rephaeli 1994, EK00),
 and may not leave a visible trace of past RG activity.
If interpreted as non-thermal emission,
 the X-ray excesses could then be the `smoking gun'
 of shock heating by RGs in the not-too-remote past for these particular groups.

A different type of non-thermal emission may also arise,
 which would be peculiar to RGs
 and related to the issue of cocoon dissipation (Sec.\ref{sec:assum})
As already mentioned,
 the exact constituents of the jet plasma inside the cocoon are unknown,
 but it is quite possible that it comprises
 primarily relativistic, non-thermal particles,
 whose energy content can be very large.
Cocoon particles of this kind injected into the ICM
 may give forth to observable non-thermal features
 distinct from the forward shock emission,
 at much later epochs.
Although such processes cannot be reliably predicted at the moment,
 this possibility needs to be explored in the future.

Finally,
 as suggested by Valageas \& Silk (1999) for active galaxies in general,
 we speculate that the large energy input into the ICM by RGs
 may have played some role in regulating the formation and evolution
 of stars and galaxies inside clusters.

\section{Conclusions}
\label{sec:conc}

We conclude by summarizing the salient points addressed in this paper.

Utilizing recently published radio LFs (W01)
 and the observed correlation between jet power and radio luminosity (Rawlings 1992, W99),
 the total energy output of RG jets in the universe
 can be evaluated with reasonable confidence,
 $U_j \sim 1-3 \times 10^{57} \rm{erg \ Mpc^{-3}}$,
 the bulk of which is attained at redshifts $z \sim 1-3$.
Combining this with the local RG LF in clusters (LO96),
 the total energy input into the ICM
 can be estimated to be 1--2 keV per particle,
 precisely in the range recent models require to reproduce
 the observed deviations from self-similarity
 in the luminosity-temperature relation of groups and clusters.
RGs have a decided edge over SNe in that
 they directly heat the ICM outside galaxies
 and are immune to radiative cooling.
The picture of ICM shock heating by RGs may be most clearly tested
 and distinguished from that due to SNe
 by observing the high energy gamma-rays
 promptly emitted by shock-accelerated electrons during the era of ICM heating,
 which can be realized by {\it GLAST}.

\acknowledgments

S.I. is grateful to a number of colleagues for valuable discussions,
 particularly T. Kitayama, as well as M. Nagashima, Y. Fujita and T. Totani.
He has also greatly benefited from conversations with E. Waxman.
We thank the anonymous referee for helpful comments.
This research was supported by the Grant-in-Aid for Scientific Research
 from the Ministry of Education, Science, Sports and Culture of Japan (No. 12304009).

\clearpage

\end{document}